\documentclass{elsarticle}

\catcode`\@=11
\def\gsim{\ifmmode{\mathrel{\mathpalette\@versim>}}
    \else{$\mathrel{\mathpalette\@versim>}$}\fi}
\def\lsim{\ifmmode{\mathrel{\mathpalette\@versim<}}
    \else{$\mathrel{\mathpalette\@versim<}$}\fi}
\def\@versim#1#2{\lower 2.9truept \vbox{\baselineskip 0pt \lineskip 
    0.5truept \ialign{$\m@th#1\hfil##\hfil$\crcr#2\crcr\sim\crcr}}}
\catcode`\@=12

\usepackage{natbib}

\usepackage{epsfig}

\usepackage{amssymb}

\begin{document}

\begin{frontmatter}

\title{Clues on black hole feedback from simulated
and observed X-ray properties of elliptical galaxies}

\author[label1]{S. Pellegrini}\ead{silvia.pellegrini@unibo.it} \author[label1]{L. Ciotti} \author[label2]{J.P. Ostriker}
\address[label1]{Astronomy Department, Bologna University, via Ranzani 1, 40127 Bologna, Italy}
\address[label2]{Princeton University Observatory, Peyton Hall, Princeton, NJ 08544, USA}

\begin{abstract} 

The centers of elliptical galaxies host supermassive black holes that
significantly affect the surrounding interstellar medium through
feedback resulting from the accretion process.  The evolution of this
gas and of the nuclear emission during the galaxies' lifetime has been
studied recently with high-resolution hydrodynamical
simulations. These included gas cooling and heating specific for an
average AGN spectral energy distribution, a radiative efficiency
declining at low mass accretion rates, and mechanical coupling between
the hot gas and AGN winds. Here we present a short summary of the
observational properties resulting from the simulations, focussing on
1) the nuclear luminosity; 2) the global luminosity and temperature of
the hot gas; 3) its temperature profile and X-ray brightness profile.
These properties are compared with those of galaxies of the local
universe, pointing out the successes of the adopted feedback and the
needs for new input in the simulations.

\end{abstract}


\end{frontmatter}

\section{Introduction}

Supermassive black holes (MBHs) at the centers of bulges and
elliptical galaxies played an important role during galaxy formation
and evolution, as revealed by remarkable correlations between their
masses and some host galaxy properties (e.g., Ferrarese \& Merritt
2000, Gebhardt et al. 2000) and by many studies
(e.g., Merloni et al. 2004; Sazonov et al. 2005; Di
Matteo, Springel \& Hernquist 2005, Hopkins et al. 2006).  An
important aspect of the coevolution process is the interaction between
the energy emitted by the accreting MBH and a galactic interstellar
medium (ISM) continuously replenished by normal stellar mass
losses. In the absence of feedback from a central MBH, this ISM
develops a flow directed towards the galactic center, accreting $\gsim
1 M_{\odot}$ yr$^{-1}$ in a process similar to a "cooling flow" (e.g.,
Ciotti et al. 1991, David et al. 1991, Pellegrini \& Ciotti 1998).
Therefore a few basic questions arise: 1) what is the fate of the
large amounts of gas accreting towards the center during the galaxies'
lifetime, and not observed at any wavelength (e.g., Peterson \& Fabian
2006)? Just $\sim $1\% of the mass made available by stars is in the
mass of present epoch MBHs (Ciotti \& Ostriker 2007).  2) how much
radiative and mechanical energy output from the MBH can effectively
interact with the surrounding ISM? 3) what are the resultant masses of
the MBHs at the present epoch?  and finally, 4) why bright AGNs are
not common in the spheroids of the local Universe, given the expected
mass accretion rate (e.g., Fabian \& Canizares 1988, Di Matteo et
al. 2003, Pellegrini 2005a)?

Recently, the interaction of the MBH with the inflowing gas has been
studied with high resolution 1D hydrodynamical simulations by Ciotti \&
Ostriker (2007), Ciotti, Ostriker \& Proga (2009, hereafter paper I)
and Ciotti \& Ostriker (2009, hereafter paper II).  These simulations
included a detailed and physically based treatment of the radiative 
energy output from the MBH and its transfer in the ISM, and of the
mechanical energy from AGN winds. They showed recurrent brief flaring
of the nucleus, with the galaxy seen alternately as an AGN/starburst
for a small fraction of the time and as a ``normal'' quiescent
elliptical for much longer intervals. Accretion feedback proved
effective in suppressing long lasting cooling flows and in maintaining
MBH masses within the range observed today (since gas is mostly lost
in outflows or starbursts); a major role in regulating the flow evolution 
was played also by type Ia supernovae (hereafter SNIa's).

Here we present a preliminary investigation of the spectral and
morphological appearance of this class of models in the X-ray band.
In particular, we focus on the hot ISM and the nucleus during the
galaxy's lifetime, both in quiescence and during outbursts of
activity, and we compare their properties with the observations
collected for galaxies of the local universe by $ROSAT$ and more
recently by $Chandra$. Section 2 briefly describes the chosen
representative model, Section 3 its observational properties and a
comparison with observational results, Section 4 summarizes
the conclusions.  We anticipate here some general results for this
class of models, and a more extensive description will be given in
Pellegrini, Ciotti \& Ostriker (2009, hereafter paper III).

\section{The representative model}\label{outbur}

The main properties of feedback modulated galactic accretion flows are
given in Ciotti \& Ostriker (2007) and recently in papers I and II,
together with a full description of the code used for the
simulations. Briefly, the code integrates the time-dependent 1D
Eulerian equations of hydrodynamics, with a logarithmically spaced and
staggered radial grid, extending from 2.5pc from the central MBH to
250 kpc.  The code calculates self-consistenly the source and sink
terms of mass, momentum and energy, associated with the evolving
stellar population (stellar mass losses and SNIa's events),
nuclear starbursts, accretion and MBH feedback.  Gas heating and
cooling are calculated for a photoionized plasma in equilibrium with
an average quasar spectral energy distribution (Sazonov et al. 2005),
the resulting radiation pressure and absorption/emission are computed
and distributed over the ISM from numerical integration of the
radiative transport equation.

Of interest here is that the adopted radiative efficiency of material
accreting on the MBH at a rate $\dot M$ is $\epsilon =0.1\times 100
\dot m/(1+100\dot m)$, where $\dot m=\dot M/\dot M_{Edd}$ is the
Eddington-scaled accretion rate, so that $\epsilon \sim 0.1$ at large
mass accretion rates (when $\dot m \gg 0.01$), and declines in a
RIAF-like fashion $\epsilon \sim 10 \dot m$ for $\dot m\lsim 0.01$ (as
for radiatively inefficient accretion flows, Narayan \& Yi 1994).  The
mechanical feedback implemented is that of quasar outflows [e.g.,
Chartas et al. (2003), Crenshaw et al. (2003)], as modelled
numerically by Proga (2003).  The mechanical efficiency scales with
the Eddington ratio ($l=L/L_{Edd})$, reaching a maximum value for
$l=2$ of $3\times 10^{-4}$ to $5\times 10^{-3}$ in different models.
Note that only the mechanical feedback due to the Broad Line Region
wind is activated (the nuclear jet properties will be inserted
in a future work).
 
Here we consider a representative model from the latest set of
simulations (paper II), for an isolated typical elliptical galaxy with
a central stellar velocity dispersion $\sigma =260$ km s$^{-1}$, a
B-band luminosity $L_B=5\times 10^{10}L_{B,\odot}$, a stellar mass
described by the Jaffe law with a mass-to-light ratio $M_*/L_B=5.8$, an
effective radius $R_e=6.9$ kpc, a dark matter halo with equal amount
of dark and visible mass within $R_e$, a standard SNIa's rate; the
maximum value of the mechanical efficiency is set to $10^{-3}$.  The
simulations begin at a galaxy age of $\sim 2$ Gyr (i.e., a redshift
$z\sim 2$, the exact value depending on the epoch of elliptical galaxy
formation, usually put at $z\gsim 2$) and soon AGN outbursts develop,
each followed by an abrupt drop of the accretion rate, and by times
during which the galaxy is replenished of gas by the stellar mass
losses.

The typical behavior of the gas during an outburst starts with the
off-center growth of a shell of denser gas (at a radius of $\sim
0.5-1$ kpc) that progressively cools, falls and reaches the center
(i.e., a cooling collapse occurs). Then a radiative shock quickly [in
$\sim (1-2) \times 10^6$ yrs] produces an outward moving shell of cold
and dense gas, that slows down until it falls back towards the center,
accumulating far more material than the first shell, and giving origin
to a larger accretion and feedback episode. After some more events of
this kind, a major shock leaves behind a very hot and dense center,
and causes substantial galaxy degassing.  The gas then cools and
resumes its subsonic velocity, the density starts increasing again and
the cycle repeats.  Overall, during outbursts both hotter and colder
regions are continuously created, the hotter ones mainly at the
center (due to gas compressed by the falling shells, or to shocks) and
the colder ones due to the gas in the falling shells or in
the radiative shocks.

We remark that the representative model has fairly standard input
parameter values, and its general properties described below are
typical of this class of models. A large set of runs exploring the
impact of different values for the feedback parameters (e.g., the wind
opening angle, the peak mechanical efficiency and its dependence on
$l$) showed present-day properties similar to those of the adopted
representative model (papers I and II).

\section{Comparison of simulated and observed X-ray properties}

We consider in turn the evolution of the luminosity of the nucleus
(Section~\ref{nuc}), of the total luminosity and 
emission weighted temperature of the hot gas (Section~\ref{global}),
of the temperature profile (Section~\ref{temp}) and of the surface
brightness profile (Section~\ref{bril}).  The emission is calculated
over the full $Chandra$ sensitivity band (0.3--8 keV), and
in two separate bands, 0.3--2 keV and 2--8
keV, by means of the APEC code within the package XSPEC for the
X-ray data analysis (see paper III for more details).

Since in the simulations the treatment of feedback is physically
based, not arbitrarily chosen and tuned to reproduce observations, any
agreement or discrepancy of the resulting model properties with
observations is relevant to improve our understanding of the MBH--ISM
coevolution.

\subsection{The nuclear luminosity}\label{nuc}

\begin{figure}
\begin{center}
\vskip -6truecm
\includegraphics[scale=0.6]{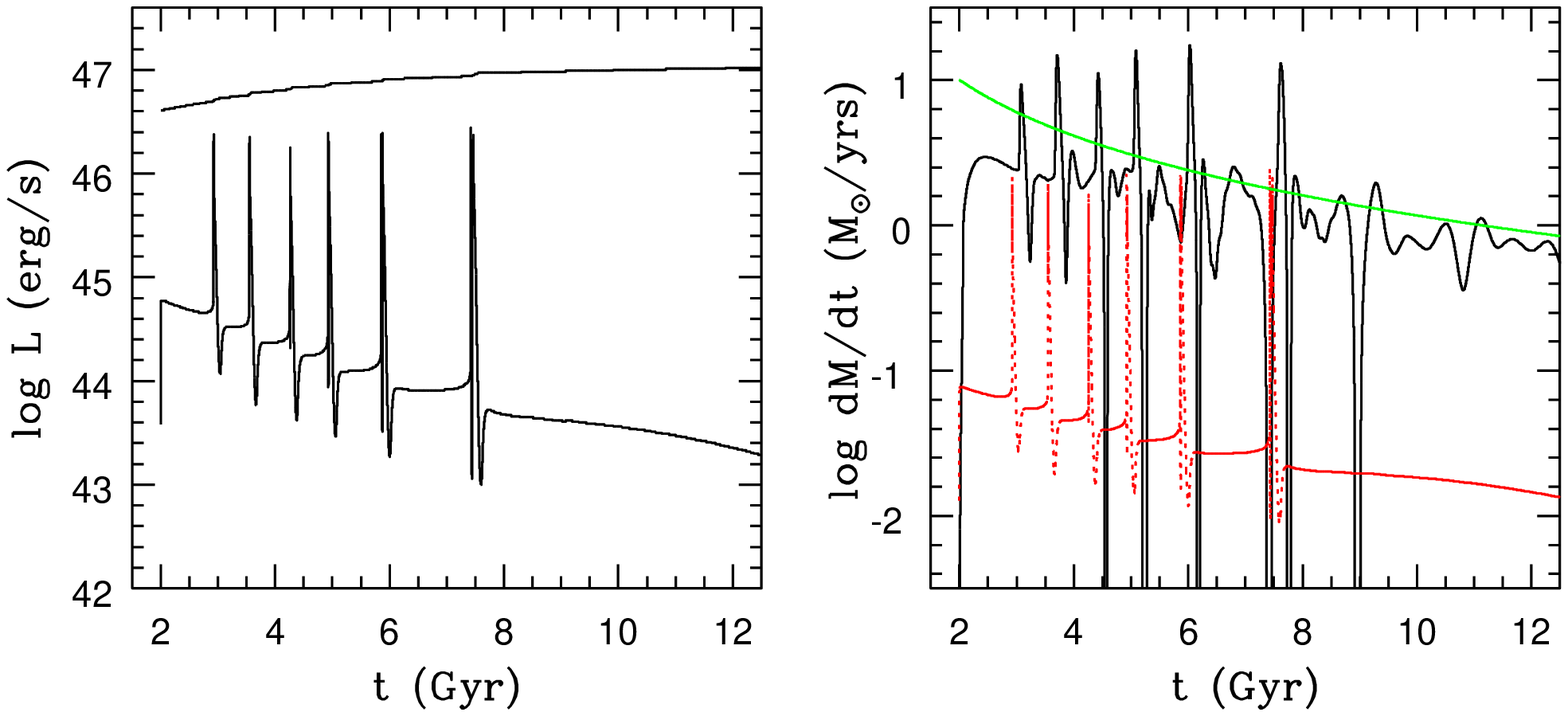}
\end{center}
\vskip -0.9truecm
\caption{
Left panel: the evolution of the bolometric luminosity produced by nuclear 
accretion. The upper line is the Eddington luminosity (note the
slow and overall  modest increase of the MBH mass). Right panel: 
the evolution of the mass budget: the stellar mass loss rate (green 
line), the gas mass leaving the galaxy at a radius of $10 R_e$ (black line) 
and the mass accretion rate on the MBH (red line).}
\label{lnuc}
\end{figure}

The time evolution of the bolometric nuclear luminosity shows strong
intermittencies at an earlier epoch, reaching the Eddington value, and
becoming rarer and rarer with time, until a smooth, very sub-Eddington
phase establishes (Fig.~\ref{lnuc}).  This behavior follows from the
secular decrease of the stellar mass loss rate (Fig.~\ref{lnuc}),
which produces longer and longer times for the replenishment of the
galaxy with gas. Towards the present epoch (a galaxy age of $\sim 12$
Gyr) the mass accretion rate is $\dot M\sim 0.01 M_{\odot}$/yr, that
is $\dot m \sim 1.1\times 10^{-3}$, therefore accretion has entered
the RIAF regime and $\epsilon \simeq 0.01$ (Sect.~\ref{outbur}).  The
nuclear bolometric luminosity is $L_{bol}=2\times 10^{43}$ erg
s$^{-1}$ and its $l=2\times 10^{-4}$, in agreement with the fact that
in the local universe the fraction of MBHs approaching their Eddington
limit is negligible.  For example, in the statistically complete
Palomar spectroscopic survey of 486 nearby galaxies only $\sim 50$\%
of ellipticals show emission line nuclei, mostly of low
level ($L_{H\alpha}<10^{40}$ erg/s); for LINERs, a sub-sample 
largely dominated by ellipticals, $L_{bol}$ ranges from $10^{39}$ to
$10^{43}$ erg s$^{-1}$, with a peak at $L_{bol}\sim 10^{41}$ erg
s$^{-1}$, and the median\footnote{$L_{bol}$ derives from the nuclear
2--10 keV emission, with $L_{bol}/L_{2-10\,keV}=16$ as appropriate for
low luminosity AGNs, and $L_{Edd}$ from MBH masses estimated from the
MBH-$\sigma$ relation.}  $l=(0.5-1)\times 10^{-5}$ (Ho
2008).  Seyferts excluded, nuclei of all types have $L_{bol} <10^{43}$
erg s$^{-1}$. Therefore the representative model may need a reduction
of its $L_{bol}$, in addition to that produced by the adopted
low radiative efficiency of RIAFs, to be more representative of the
typical low luminosity nucleus of the local universe.

Indeed, many nuclei of nearby ellipticals studied in detail in the
X-ray band show emission at an extremely low level. 
Their 0.3--10 keV luminosities $L_{X,nuc}$, for a sample
within $\sim 50$ Mpc, with or without radio emission and
residing in all kinds of environments, range from
few$\times 10^{38}$ erg s$^{-1}$ to $\sim 10^{42}$ erg s$^{-1}$, with
$L_{X,nuc}/L_{Edd}$ as low as $10^{-6}-10^{-8}$ the most common
(Pellegrini 2005a,b; see also Soria et al. 2006; Gallo et
al. 2008). $L_{X,nuc}$ of the model at the present epoch can be
derived from $L_{bol}$ adopting a correction factor appropriate for
the spectral energy distribution of a RIAF, that is $\lsim 0.2$ 
for low luminosity MBHs (Mahadevan 1997). This gives $L_{X,nuc} \lsim 4\times
10^{42}$ erg s$^{-1}$, and $L_{X,nuc}/L_{Edd}\lsim 4\times
10^{-5}$, both values again quite higher than typically observed.

A reduction of the model nuclear luminosity could be accomplished by
inserting the mechanical feedback\footnote{We recall that the
simulations include only the mechanical feedback due to the Broad Line
Region wind (see Sect.~\ref{outbur} and Fig. 1 in paper I).} of a jet
or of a nuclear wind, that should naturally develop in low radiative
efficiency accretion (Blandford \& Begelman 1999), and should reduce
further the mass accretion rate (e.g., Di Matteo et al. 2003).  In
fact, the Bondi accretion rate for the hot ISM around many nearby
MBHs, coupled to the RIAF radiative efficiency, corresponds to
luminosities still higher than observed, indicating the need for a
reduction of the mass that actually reaches the MBH (Loewenstein et
al. 2001, Pellegrini 2005a, Pellegrini et al. 2007a,b). Another
indication of the relevance of mechanical feedback is the observed
link between the radio luminosity and the X-ray gas asymmetry index
for $\sim 50$ nearby ellipticals (Diehl \& Statler 2008a).

Finally, note that $L_{X,nux}$ is observed to be independent of the
MBH mass and of the Bondi accretion rate (Pellegrini
2005a).  Likely, neither Bondi accretion nor simple RIAFs
realistically describe the interaction between the outgoing energy
flux and the incoming mass flux at the nucleus, while
the fluctuations typical of feedback cycles can 
account for the observed lack of relationship.

\subsection{Global gas luminosity and temperature}\label{global}

\begin{figure}
\begin{center}
\vskip -2truecm
\includegraphics[width=10cm]{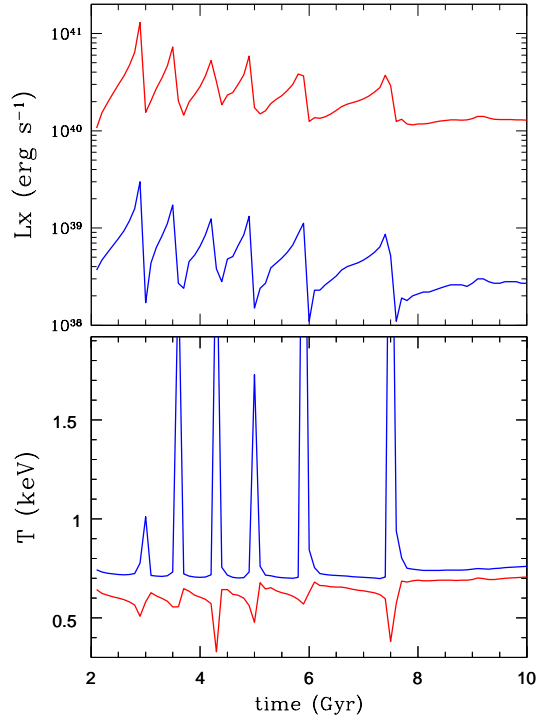}
\end{center}
\vskip -0.5truecm
\caption{Time evolution of the total gas luminosity and emission weighted 
temperature within the optical effective radius $R_e$, in the 0.3--2 keV band
(red lines) and in  the 2--8 keV band (blue lines).}
\label{fig1}
\end{figure}

The time evolution of the gas emission shows peaks during outbursts,
that become broader and less high with time increasing
(Fig.~\ref{fig1}).  Soft and hard emission oscillate in phase and
present the same overall behavior, with the hard one keeping $\lsim
100$ times lower.  Hard emission during outbursts would be difficult
to distinguish from the contribution of unresolved binaries even with
$Chandra$, if extended (e.g., Trinchieri et al. 2008), but it could
be detected if centrally concentrated (see also Sect.~\ref{bril}).

The largest compilation of the global X-ray emission of early type
galaxies of the local universe is based on $ROSAT$ observations of a
relatively unbiased sample of 401 objects, homogeneously analyzed
(O'Sullivan et al. 2001).  The best fit correlation with the B-band
luminosity of the emission mostly due to hot gas ($L_{X,ISM}$), after
subtraction of the stellar contribution, predicts $L_{X,ISM} \sim
10^{41}$ erg s$^{-1}$ for $L_B=5\times 10^{10}L_{B,\odot}$ as the
model galaxy. However at this $L_B$, due to the large scatter around
the best fit, the observed $L_{X,ISM}$ ranges from $10^{40}$ erg
s$^{-1}$ up to few $\times 10^{42}$ erg s$^{-1}$ (though $L_{X,ISM}>$
few$\times 10^{41}$ erg s$^{-1}$ belong to galaxies with an important
contribution from the intragroup or intracluster medium).  At an age
of $\sim 10-12$ Gyr the model has $L_{X,ISM}\sim 10^{40}$ erg s$^{-1}$
(Fig.~\ref{fig1}), on the lower end of the observed range.  This
indicates that degassing is too efficient in the simulations, at least
at the present epoch. This problem would be alleviated by 
an external pressure from an outer medium (e.g., Vedder
et al. 1988), a more appropriate "boundary" in fact for
ellipticals that usually are not isolated (it will be considered in
future simulations).

The lower panel of Fig.~\ref{fig1} shows the evolution of the soft
($<T>_s$) and hard ($<T>_h$) emission weighted temperatures within the
optical $R_e$, and their complex behavior during outbursts. The sharp
and high peaks in $<T>_h$ correspond to the onset of very hot regions,
while the decrements in $<T>_s$ are due to a dense cold shell
preceeding the major burst and to cold gas swept by the radiative shocks
produced by the outburst (see Sect.~\ref{outbur}).  The 0.3--8
keV emission weighted temperature (not shown in Fig.~\ref{fig1}) is
almost coincident with $<T>_s$, except for those short times with a
very hot gas component.

$ROSAT$ observations showed a correlation, though with a large degree
of scatter, between the temperature weighted with the soft 
emission of the whole galaxy ($T$) and the central $\sigma$
(O'Sullivan et al. 2003), predicting $T\sim 0.7$ keV (with observed
values from $\sim 0.4$ to $\sim 1$ keV) for $\sigma =260$ km s$^{-1}$
as for the model galaxy.  At the present epoch the model has $T \sim
0.5$ keV, lower than the best fit prediction but within the observed
range, and $<T>_s=0.7$ keV (Fig.~\ref{fig1}). Likely the observed $T$
refers to a region less extended radially (and typically also hotter)
than used for the model galaxy.  Average temperatures within $R_e$
have been obtained with $Chandra$ (Athey 2007), whose high angular
resolution allows for a better subtraction of the point (stellar)
source contribution and then gas temperatures closer to the true ones.
For 22 Athey's galaxies with log $L_B$=10.5 to 10.8, similar to that
of the model, the average 0.3--8 keV emission weighted temperature is
$0.60$ keV, close to $<T>_s$ though slightly lower.

Summarizing, the gas content of the model is lower than the average
observed for a galaxy of its size; this shows the need for a lower
SNIa heating, or a higher/more concentrated gravitating mass, or the
confining effect of an external medium.  The average temperatures instead
agree reasonably with observations, and modifications to the model
should not increase the average temperature within $R_e$, while they
could increase a little that over the whole galaxy.

\subsection{Temperature profiles}\label{temp}

\begin{figure}
\vskip -2.5truecm
\hskip 1.5truecm
\includegraphics[angle=0.,scale=0.5]{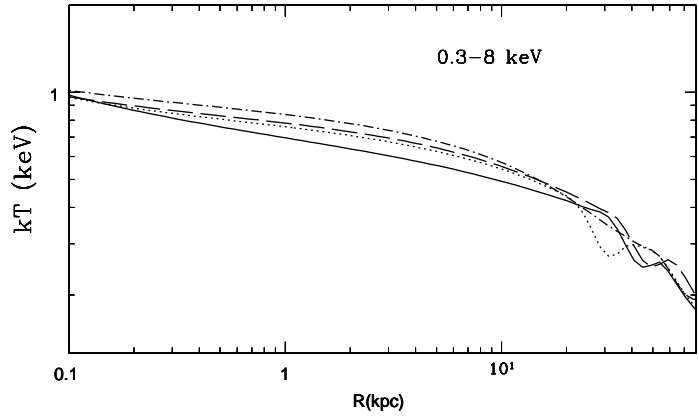}
\vskip -4.5truecm
\caption{Projected temperature profiles weighted with the emission in
the 0.3-8 keV band, during the interburst times
t=6.9 Gyr (solid line), 8.1 Gyr (dotted), 9.1 Gyr (dashed) and 
11.1 Gyr (dot-dashed line).}  
\label{t2}
\end{figure}

\begin{figure}
\vskip -1.5truecm
\hskip 1.5truecm
\includegraphics[scale=0.5]{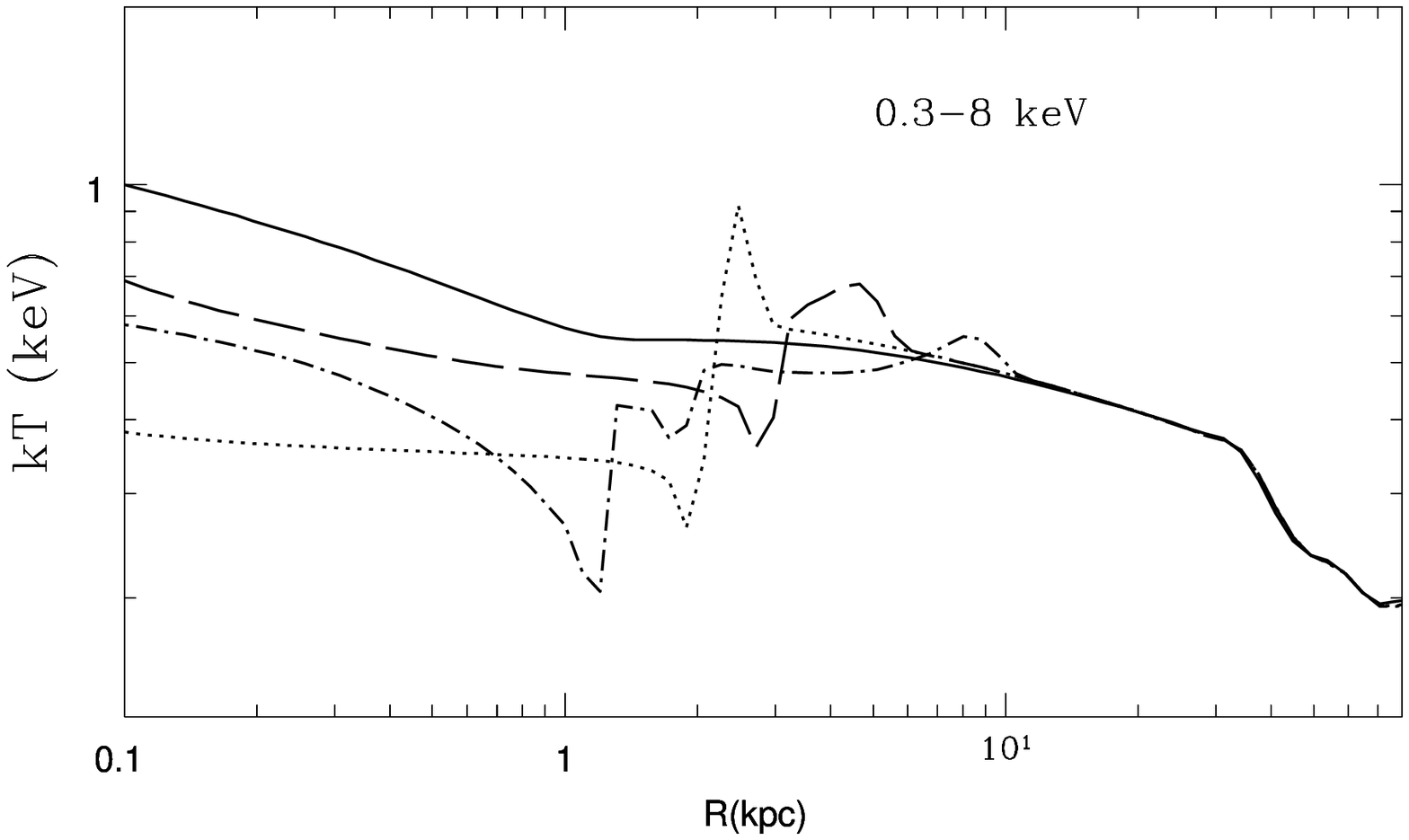}
\vskip -4.5truecm
\hskip 1.5truecm
\includegraphics[scale=0.5]{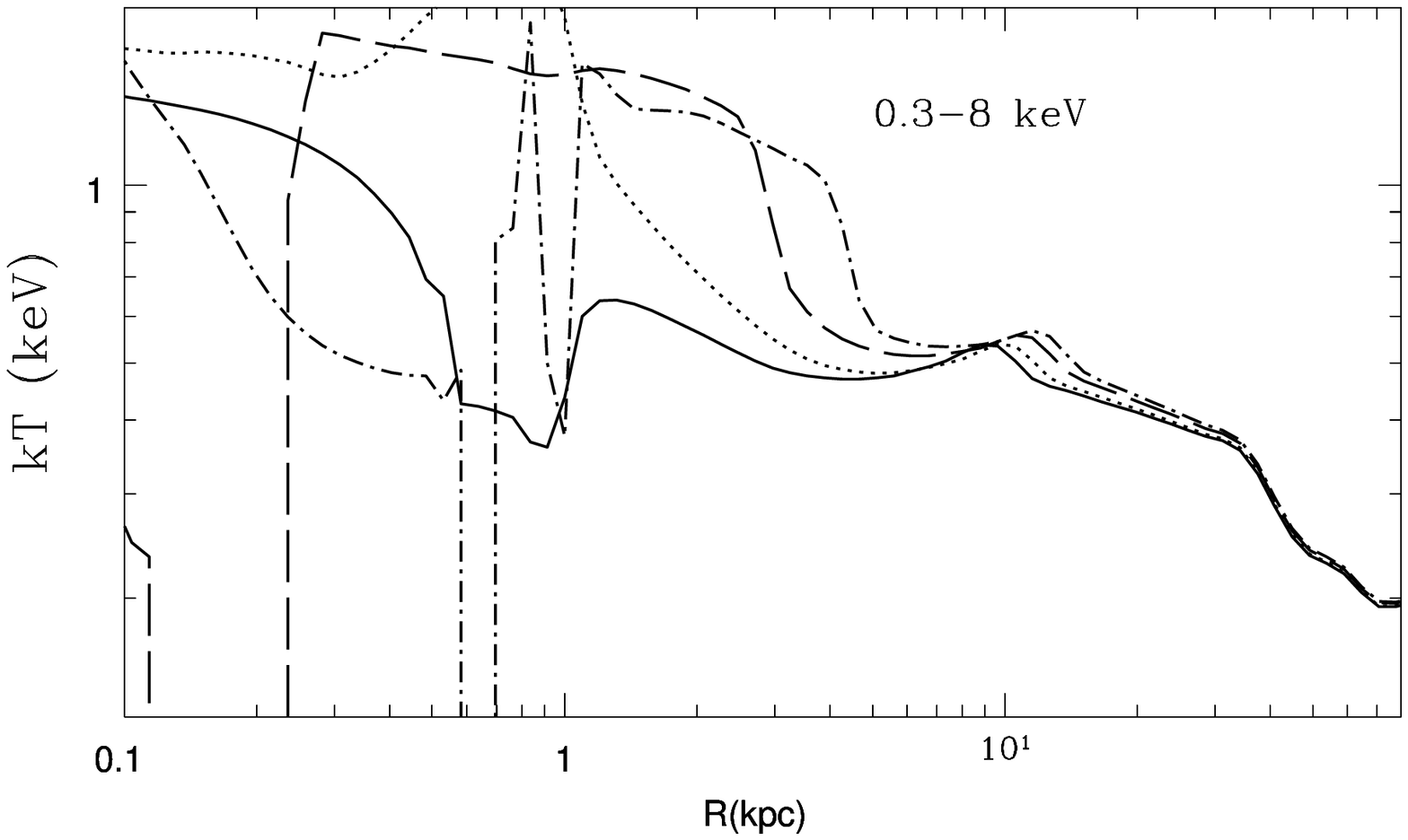}
\vskip -4.5truecm
\hskip 1.5truecm
\includegraphics[scale=0.5]{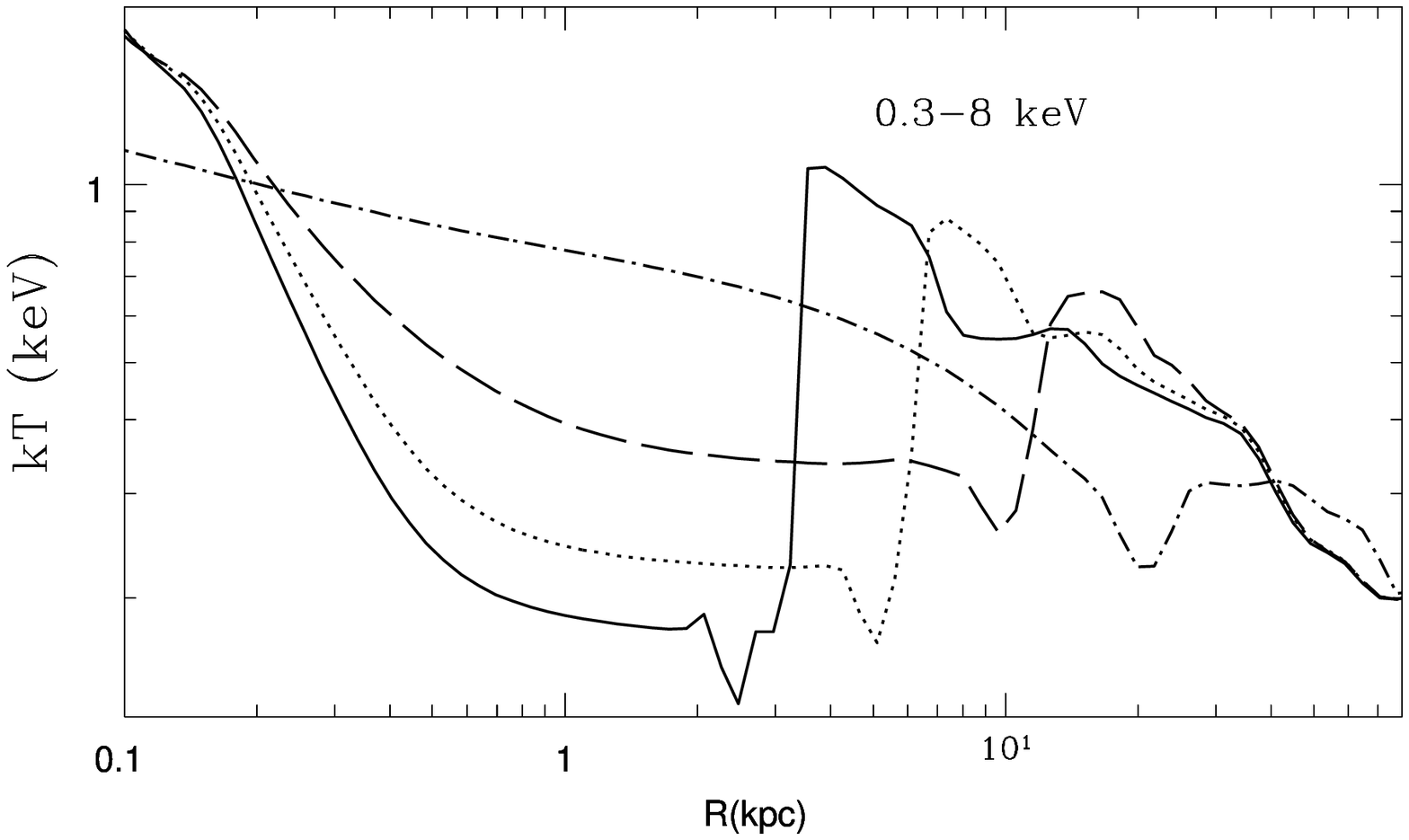}
\vskip -4.5truecm
\caption{Projected temperature profiles for the gas emission during
the last outburst (see Sect.~\ref{outbur} for more details).  Upper panel: the shell is forming at a radius of
$\sim 1$ kpc and approaching the center (solid line); a shock is
moving outward after the first burst (dotted line, after 54 Myr; 
dashed line, after 4 Myr), creating a
major shell that falls back to the center (dot-dashed line, after
8 Myr). Central panel (the time step between each line is
2 Myr): after the shell has reached the center, a
shock is moving outward, creating a central very hot bubble (solid
line, dotted line), that quickly cools (dashed and dot-dashed
line). Bottom panel: the shock 
is moving outwards [solid ($+4$ Myr), dotted ($+6$ Myr), 
dashed ($+16$ Myr) lines]; quiescence is resuming
(dot-dashed line, $+0.1$ Gyr). The time elapsed from the beginning of
the outburst (Fig.~\ref{t3}, upper panel, solid line) is 0.2
Gyr. }
\label{t3} 
\end{figure}

The emission-weighted and projected temperature profiles are shown
during quiescent inter-burst periods in Fig.~\ref{t2}, and during the
last outburst at $\sim 7.5$ Gyr in Fig.~\ref{t3}.  During quiescent
times the profiles are smooth, with the temperature monothonically
decreasing for increasing radius (from $\sim 1$ keV at a radius of
$\sim 100$ pc to $\sim 0.4-0.5$ keV at $\sim 20$ kpc).  Negative
temperature gradients are typical of inflowing gas in steep potentials
(e.g., Pellegrini \& Ciotti 1998), just due to compressional heating.
Diehl \& Statler (2008b) suggest that negative temperature gradients
could be the sign of localized heating by a central, weak AGN. In fact
in the models, even during quiescent times, accretion is present
(though with a low radiative efficiency, Sects.~\ref{outbur} and
\ref{nuc}), and this has an additional heating effect over
gravitational compression on the gas at the center.  Negative
gradients are commonly observed among ellipticals, as revealed by
$Chandra$ for an increasing number of galaxies (Fukazawa et al. 2006,
Athey 2007, Diehl \& Statler 2008b), and the temperature roughly
halves from the center to the galaxy outskirts, as in the model. A few
galaxies have central temperatures of $\sim 1$ keV (as in
Fig.~\ref{t2}), while most show temperatures between 0.5 and 1 keV at
their innermost radial bin. However, the model temperature calculated
as an average for a central bin extending out to 0.5--a few kpc, as
for observed galaxies, will be lower than 1 keV. While a more detailed
comparison is postponed to paper III, we conclude here that no
additional heating seems to be required in the central region (as
found in Sect.~\ref{global}).

In other observed profiles the temperature increases outward, or keeps
roughly flat, or is "hybrid", firstly decreasing from the central
value to a minimum and then increasing out to the edge of the galaxy
(Diehl \& Statler 2008b).  These profiles could be produced in the
model by environmental effects currently not included 
(e.g., Vedder et al. 1988, Mathews \& Brighenti 2003). A temperature decreasing
towards the center is also observed, but is never shown by the model
during quiescence.  A central drop in temperature and a shape
resembling the hybrid profile are shown during particular phases of
outbursts, when the major shock is moving outwards or the cold shell
is forming (see Fig.~\ref{t3} that illustrates the messy
temperature distribution during the $\sim 0.2$ Gyr of an
outburst, consequent to the gas flow behavior described in
Sect.~\ref{outbur}).

We noted (Sect.~\ref{nuc}) that the stationary hot accretion phase
needs a reduction of the accretion rate, as can be provided by a jet.
In a recent $Chandra$ analysis the temperature profiles outside $\sim
R_e$ showed gradients that switch monothonically from positive to
negative going from high to low radio luminosities within three $R_e$
(Diehl \& Statler 2008b).  To be consistent with this result, the jet
addition to the simulations should heat the gas outside $\sim R_e$.

Finally note that the temperature profile of the elliptical NGC3411
shows an intriguing off-center dip (O'Sullivan et al. 2007) that is
highly unusual for standard cooling flow models but is present in a few
profiles in Fig.~\ref{t3}. This dip corresponds to a ripple in the
observed X-ray brightness profile, that is also a frequent feature in
the model (Sect.~\ref{bril}).

\subsection{Brightness profiles}\label{bril}

\begin{figure}
\begin{center}
\vskip -6truecm
\includegraphics[scale=0.6]{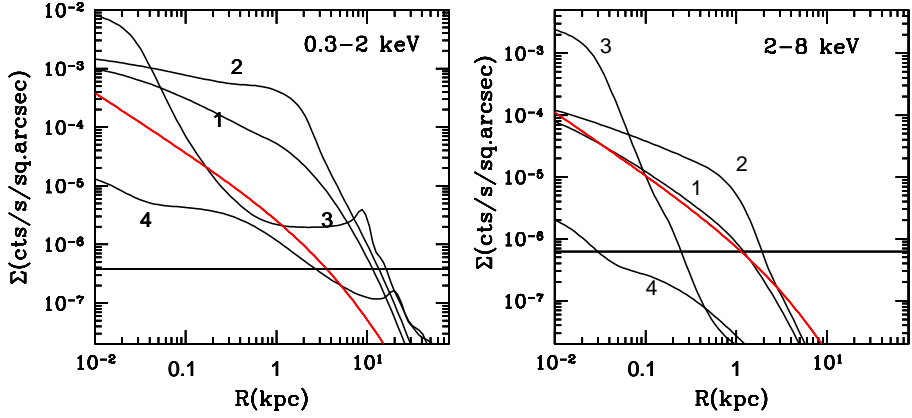}
\end{center}
\vskip -0.6truecm
\caption{The brightness profile during the last major outburst; 
curves referring to subsequent
times are labeled from 1 to 4 (see Sect.~\ref{bril} for a description
of the main features).
Count rates in the y-axis and background level (straight horizontal
line) are for the $Chandra$ ACIS-S detector. 
The red line shows the profile of unresolved X-ray
binaries, appropriate for their spectral shape and a representative
observation of a Virgo galaxy with an exposure of 200 ksec (e.g., Kim
et al. 2006).}
\label{brilou}
\end{figure}

\begin{figure}
\begin{center}
\vskip -1.truecm
\includegraphics[scale=0.5]{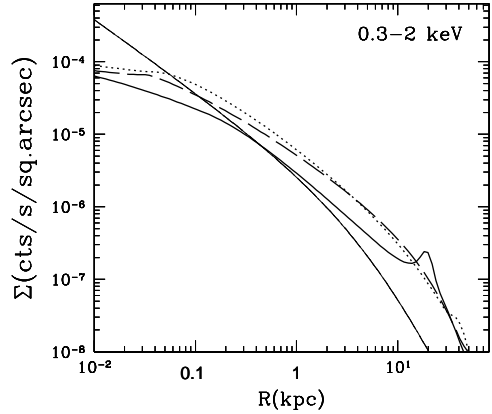}
\end{center}
\vskip -5.truecm
\caption{The evolution of the brightness profile after the last major
outburst: the quiescent state
establishes again and is kept until the end of the simulation.  The
solid, dotted, dashed lines correspond to times of 7.7, 8.7 and 10.0 Gyr.
Count rates and unresolved stellar emission profile (thin smoother line)
have been calculated as for Fig.~\ref{brilou}.}
\label{brilfin}
\end{figure}

The evolution of the surface brightness profile is shown in
Fig.~\ref{brilou}, right before and during the last major outburst,
and in Fig.~\ref{brilfin}, during the subsequent slow evolution until
the present epoch quiescent state. In Fig.~\ref{brilou} curves
labelled 1 and 2 show two subsequent times of the off-center shell
formation, lasting $\sim 0.5$ Gyr.  Curve labelled 3 shows a shock
moving outwards after the major outburst (in the upper panel) and the
presence of very hot gas at the center (lower panel); this phase lasts
for $\sim 2\times 10^7$ yr. Curve labelled 4 shows the result of the
degassing caused by the passage of the shock waves: the galaxy has a
low gas density and subsonic perturbations remain at a radius of a few
tens of kpc; the time is $\lsim 0.2$ Gyr since the outburst started.
Note how the main features in the profile are above the levels of
background and of unresolved stellar emission (described in the
caption of the figures).
 The high-temperature, high-density phase at the center (curve 3)
is very brief, and unlikely to be observed. Disturbances as shells and
ripples lasting $\lsim 0.2$ Gyr are more likely to be observed. In
fact, many nearby galaxies show these characteristics in their
images, as recently proven with $Chandra$ studies (Diehl \& Statler
2008a, Machacek et al. 2006).

Interestingly, three bright ellipticals imaged with $Chandra$
(Loewenstein et al. 2001) all show a flattish X-ray brightness
profile within the central 3--10 arcsec ($\lsim 1$ kpc), that is not
possible to reproduce with inflow models, while it resembles the
profile of the "pre-burst"phase (curves 1 and 2 in Fig.~\ref{brilou}).

\section{Summary and conclusions}

We have calculated the X-ray properties of a galaxy model
representative of a recent investigation of feedback modulated
accretion including the combined heating of radiation and AGN winds
(papers I and II), and we have compared them with observations of the local
universe. Besides important achievements as limiting the growth of
the MBH mass to observed values (papers I and II), thanks to the
present analysis these models also reveal important aspects of 
the MBH--ISM coevolution, as detailed below.

At the present epoch a highly sub-Eddington ($l \sim 10^{-4}$) phase
establishes with accretion in the low radiative efficiency regime.
Though within the range observed, the nuclear emission
is quite larger than the most frequently observed values [$l\sim
(0.5-1)10^{-5}$].  This suggests that an additional mechanism reducing
further the mass available for accretion is important, as could be
provided by a nuclear jet or wind from a RIAF.

During the many outbursts, the gas emission $L_{X,ISM}$ presents peaks
that become broader as time increases; sharp peaks and decrements are
shown respectively by the hard and soft emission weighted temperature.
At the final quiescent times, $L_{X,ISM}$ is on the lower side of
those observed, which suggests degassing may be less efficient, or the
confining agents (gravitating mass, external medium) more
efficient. The emission weighted temperatures instead compare well
with observations.

The profiles of brightness and temperature (decreasing outward) in
quiescence resemble those of many local galaxies. Outbursts produce
features in the brightness profile that are detectable with
$Chandra$, last for $\lsim 0.2$ Gyr, and could match part of the
widespread gas disturbances observed in local galaxies.
The hot bubbles inflated at the center produce central peaks in the 
hard band brightness profile, but their short duration makes them 
difficult to observe.

In conclusion, a form of feedback lowering further the nuclear
luminosity during the stationary hot accretion phase seems important.
This should not increase the central temperature that is already consistent
with observed values.  A jet interacting mainly outside $\sim 0.5-1
R_e$ would then be preferable; it should also heat the gas there, to
bring the overall temperature profile in accord with those observed
(increasing outward) at high radio luminosities.  The additional
feedback should not lower $L_{X,ISM}$.

Finally, we comment on how $L_{X,ISM}$ and the gas temperature might
scale with galaxy properties, and with how much scatter.  A change of
the mass content and distribution, SNIa rate and external pressure will
have an impact on the gas content, and also on the possibility of
having outbursts until the present epoch, or no outbursts at all.  For
example, the SNIa heating may keep small galaxies in an outflow so
that outbursts never happen; due to a larger dark matter content,
galaxies as the representative model may never experience a huge
degassing, and may host nuclear outbursts until the present epoch.  In
general, we expect a trend of $L_{X,ISM}$ with $L_B$ mainly determined
by the galaxy structure and SNIa's heating, as described by older
studies (e.g., Pellegrini \& Ciotti 1998), to which scatter is added
by feedback, but only for luminous galaxies (as the representative
model).  The frequency of outbursts at the present epoch is difficult
to estimate, given the many parameters involved, but it should be very
small (papers I and II).  The hot gas disturbances shown by $Chandra$
have been mostly attributed to an ongoing or recent jet activity
(e.g., Diehl \& Statler 2008a,b), though evidence for this is not
always present (e.g., NGC4552, Machacek et al. 2006).  A statistically
complete sample of galaxies with information on the nature of activity
and of the gas disturbances will hopefully be built soon from the
large $Chandra$ database.

\end{document}